\documentclass[%
 twocolumn,
 superscriptaddress,
 nofootinbib,
 amsmath,amssymb,
 aps,
 prl,
 10pt
]{revtex4-2}

\usepackage{graphicx}
\usepackage{amsmath}
\usepackage{siunitx}
\usepackage{xcolor}
\usepackage{dcolumn}
\usepackage{bm}
\usepackage{tabularx}
\usepackage{soul}

\begin{document}

\title{Single-particle incoherent diffractive imaging and amplified spontaneous emission in copper nanocubes}

\author{Tamme Wollweber}
\affiliation{Max Planck Institute for the Structure and Dynamics of Matter, 22761 Hamburg, Germany}
\affiliation{The Hamburg Centre for Ultrafast Imaging, 22761 Hamburg, Germany}
\author{Sarodi Jonak Dutta}
\affiliation{The Hamburg Centre for Ultrafast Imaging, 22761 Hamburg, Germany}
\affiliation{Department of Physics, Universität Hamburg, 22761 Hamburg, Germany}
\author{Zhou Shen}
\affiliation{Max Planck Institute for the Structure and Dynamics of Matter, 22761 Hamburg, Germany}
\author{Johan Bielecki}
\affiliation{European XFEL, 22869 Schenefeld, Germany}
\author{Carl Caleman}
\affiliation{Department of Physics and Astronomy, Uppsala University, Uppsala SE-75120, Sweden}
\affiliation{Center for Free-Electron Laser Science CFEL, Deutsches Elektronen-Synchrotron DESY, 22607 Hamburg, Germany}
\author{Sebastian Cardoch}
\affiliation{Department of Physics and Astronomy, Uppsala University, Uppsala SE-75120, Sweden}
\author{Armando D. Estillore}
\affiliation{Center for Free-Electron Laser Science CFEL, Deutsches Elektronen-Synchrotron DESY, 22607 Hamburg, Germany}
\author{Lukas V. Haas}
\affiliation{The Hamburg Centre for Ultrafast Imaging, 22761 Hamburg, Germany}
\affiliation{Department of Physics, Universität Hamburg, 22761 Hamburg, Germany}
\affiliation{Center for Free-Electron Laser Science CFEL, Deutsches Elektronen-Synchrotron DESY, 22607 Hamburg, Germany}
\author{Sebastian Karl}
\affiliation{Department of Physics, Friedrich-Alexander-Universität Erlangen-Nürnberg, 91058 Erlangen, Germany}
\author{Faisal H.M. Koua}
\affiliation{European XFEL, 22869 Schenefeld, Germany}
\author{Abhishek Mall}
\affiliation{Max Planck Institute for the Structure and Dynamics of Matter, 22761 Hamburg, Germany}
\author{Parichita Mazumder}
\affiliation{Max Planck Institute for the Structure and Dynamics of Matter, 22761 Hamburg, Germany}
\author{Diogo Melo}
\affiliation{European XFEL, 22869 Schenefeld, Germany}
\author{Mauro Prasciolu}
\affiliation{Center for Free-Electron Laser Science CFEL, Deutsches Elektronen-Synchrotron DESY, 22607 Hamburg, Germany}
\author{Omkar V. Rambadey}
\affiliation{Max Planck Institute for the Structure and Dynamics of Matter, 22761 Hamburg, Germany}
\author{Amit Kumar Samanta}
\affiliation{The Hamburg Centre for Ultrafast Imaging, 22761 Hamburg, Germany}
\affiliation{Center for Free-Electron Laser Science CFEL, Deutsches Elektronen-Synchrotron DESY, 22607 Hamburg, Germany}
\author{Abhisakh Sarma}
\affiliation{European XFEL, 22869 Schenefeld, Germany}
\author{Tokushi Sato}
\affiliation{European XFEL, 22869 Schenefeld, Germany}
\author{Egor Sobolev}
\affiliation{European XFEL, 22869 Schenefeld, Germany}
\author{Fabian Trost}
\affiliation{European XFEL, 22869 Schenefeld, Germany}
\author{Saša Bajt}
\affiliation{The Hamburg Centre for Ultrafast Imaging, 22761 Hamburg, Germany}
\affiliation{Center for Free-Electron Laser Science CFEL, Deutsches Elektronen-Synchrotron DESY, 22607 Hamburg, Germany}
\author{Richard Bean}
\affiliation{European XFEL, 22869 Schenefeld, Germany}
\author{Jochen K{\"u}pper}
\affiliation{The Hamburg Centre for Ultrafast Imaging, 22761 Hamburg, Germany}
\affiliation{Department of Physics, Universität Hamburg, 22761 Hamburg, Germany}
\affiliation{Center for Free-Electron Laser Science CFEL, Deutsches Elektronen-Synchrotron DESY, 22607 Hamburg, Germany}
\affiliation{Department of Chemistry, Universit{\"a}t Hamburg, 20146 Hamburg, Germany}
\author{Nicusor Timneanu}
\affiliation{Department of Physics and Astronomy, Uppsala University, Uppsala SE-75120, Sweden}
\author{Ralf R{\"o}hlsberger}
\affiliation{The Hamburg Centre for Ultrafast Imaging, 22761 Hamburg, Germany}
\affiliation{Deutsches Elektronen-Synchrotron DESY, 22607 Hamburg, Germany}
\affiliation{Helmholtz-Institut Jena, 07743 Jena, Germany}
\affiliation{GSI Helmholtzzentrum für Schwerionenforschung, 62491 Jena, Germany}
\affiliation{Institut für Optik und Quantenelektronik, Friedrich-Schiller-Universität Jena, 07743 Jena, Germany}
\author{Joachim von Zanthier}
\affiliation{Department of Physics, Friedrich-Alexander-Universität Erlangen-Nürnberg, Staudtstrasse 1, 91058 Erlangen, Germany}
\affiliation{Erlangen Graduate School in Advanced Optical Technologies (SAOT), Friedrich-Alexander Universität Erlangen-Nürnberg, 91052 Erlangen, Germany}
\author{Florian Schulz}
\affiliation{The Hamburg Centre for Ultrafast Imaging, 22761 Hamburg, Germany}
\affiliation{Department of Physics, Universität Hamburg, 22761 Hamburg, Germany}
\author{Henry N. Chapman}
\affiliation{The Hamburg Centre for Ultrafast Imaging, 22761 Hamburg, Germany}
\affiliation{Department of Physics, Universität Hamburg, 22761 Hamburg, Germany}
\affiliation{Center for Free-Electron Laser Science CFEL, Deutsches Elektronen-Synchrotron DESY, 22607 Hamburg, Germany}
\author{Kartik Ayyer}
\email{kartik.ayyer@mpsd.mpg.de}
\affiliation{Max Planck Institute for the Structure and Dynamics of Matter, 22761 Hamburg, Germany}
\affiliation{The Hamburg Centre for Ultrafast Imaging, 22761 Hamburg, Germany}

\begin{abstract}
We demonstrate element-specific incoherent diffractive imaging (IDI) of single copper nanocubes using intensity correlations of K$\alpha$ fluorescence at a hard X-ray free-electron laser. Combining single-particle diffraction classification with IDI, we retrieve the form factor of \SI{88}{\nano\meter} cubes with \SI{20}{\nano\meter} resolution, extending IDI to the destructive single-particle regime with a large gain in resolution. IDI visibility drops sharply above a fluence of \SI[per-mode=symbol]{E2}{\joule\per\centi\meter\squared}, consistent with the assumption of amplified spontaneous emission. Our results reveal fundamental limits for high-fluence nanoimaging towards future single-particle X-ray imaging.
\end{abstract}

\maketitle

X-ray fluorescence has been a mainstay of elemental identification since the first characteristic emission lines were identified. When combined with a scanning probe, one can spatially map the distribution of elements using so-called X-ray fluorescence microscopy (XFM). To achieve resolutions better than the focal spot size, one would like to use lensless imaging methods like coherent diffractive imaging (CDI). Unfortunately, the emission of fluorescence results in random initial phases which do not produce static interference patterns in the far field~\cite{slowik2014incoherent}, yielding instead a uniform integrated intensity distribution with no structural information. However, when fluorescence is detected within its coherence time $\tau_c$, the period during which the relative phases are stable, stationary interference patterns can be extracted using second-order intensity correlations~\cite{classen2017incoherent}, as first identified by Hanbury Brown \& Twiss (HBT) in the context of stellar imaging~\cite{brown1956test}.

More precisely, we calculate the second-order degree of coherence $g^{(2)}(\mathbf{q})$ of the fluorescence,
\begin{equation}
  g^{(2)}(\mathbf{q}) = \frac{\left\langle I (\mathbf{k}) I(\mathbf{k} + \mathbf{q}) \right\rangle_\mathbf{k}}{\left\langle I(\mathbf{k}) \right\rangle_\mathbf{k}^2} 
  \label{eq:g2_def}
\end{equation}
which is related to the Fourier transform of the spatial distribution of emitters, $\mathcal{F}(\mathbf{q})$, by the well-known Siegert-relation~\cite{goodman2015statistical},
\begin{equation}
  g^{(2)}(\mathbf{q}) = 1 + \nu \frac{\left| \mathcal{F}(\mathbf{q}) \right |^2}{\left | \mathcal{F}(0) \right | ^ 2} \label{eq:g2_siegert}
\end{equation}
where $\mathbf{q}$ is the wave vector difference between photons recorded on two detector pixels and $\nu$ is the visibility factor determining the contrast of the correlation function at $\mathbf{q}=0$. 

\begin{figure*}[ht]
  \centering
  \includegraphics[width=\textwidth]{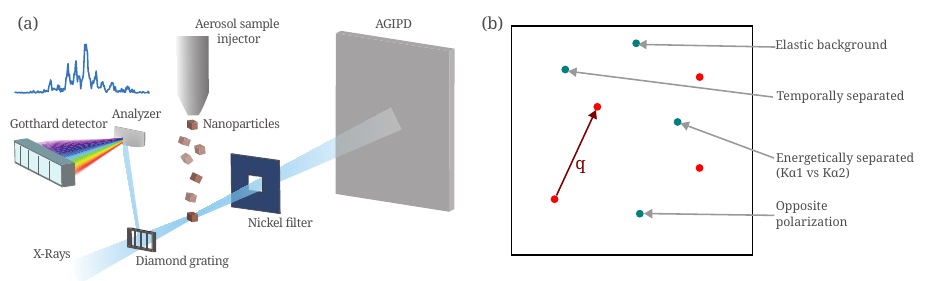}
  \caption{(a) Experimental schematic showing X-ray pulses incident on cubic copper oxide nanoparticles (not to scale) which were aerosolized and injected into the interaction region. A representative spectrum, which was measured for every pulse, is shown on the top left. The nickel foil filters the elastically scattered photons at high scattering angles, while letting the Cu-K$\alpha$ photons through. The coherent diffraction going through the Ni aperture was used for orientation determination and classification, while the fluorescence was studied in the filtered area of the detector. (b) Illustration of a region of the detector with individually measured photons. The wave-vector difference $\mathbf{q}$ as well as non-interfering photons are shown.}
  \label{fig:setup}
\end{figure*}

Incoherent diffractive imaging (IDI) bridges a critical resolution gap between local probes, such as Extended X-ray Absorption Fine Structure (EXAFS), and lower-resolution XFM techniques. Core-level fluorescence offers exceptionally high elemental sensitivity, enabling dark-field imaging of the substructure of a single element within a heterostructure. As we show below, intensity correlation analysis also provides a unique probe into the degree of coherence of the emitted fluorescence, which is especially interesting for highly intense and short XFEL pulses with non-trivial X-ray matter interactions.

However, the signal-to-noise ratio (SNR) requirements for IDI are stringent~\cite{trost2020photon,lohse2021incoherent}, depending on factors such as the number of detected photons per frame, the number of modes in the emitted light field, and the dimensions and geometry of the emitting object. These requirements for sample structure and data volume pose significant challenges for applying IDI across diverse systems. In pioneering proof-of-principle experiments, copper K$\alpha$ fluorescence was used to determine the focus profile and the pulse duration of an XFEL pulse~\cite{nakamura2020focus,inoue2019determination}. This was followed by the imaging of a non-trivial 2D beam profile using this approach~\cite{trost2023imaging}. Parallel theoretical work on the feasibility of IDI~\cite{ho2021fluorescence,shevchuk2021imaging}, including the use of higher-order correlations~\cite{peard2023ab,bojer2024phase}, and recent demonstrations of 3D imaging using fluorescence intensity correlations in stationary bulk samples~\cite{radloff2025fluorescence} further underscore the growing scope of IDI-style methods.

A major unresolved question in the field is whether IDI could be applied to single particles, and thus be used for element-specific imaging of nanoparticles~\cite{lohse2021incoherent}. The fluorescence output of a single particle is orders of magnitude lower than that of bulk samples, such as \SI{10}{\micro\meter}-scale copper foils, which have been the primary focus of earlier studies~\cite{inoue2019determination,trost2023imaging}.

In this paper, we demonstrate imaging beyond focal spot characterization by analyzing the emitted fluorescence from thousands of cubic copper nanoparticles exposed to XFEL pulses one at a time. The particles were destroyed after a single exposure so the samples were constantly refreshed. Simultaneously, single particle coherent diffractive imaging techniques were used to classify particles by size and orientation. The averaged IDI intensity correlation shows imaging to a resolution of \SI{20}{\nano\meter}, but with lower visibility, $\nu$, than predicted. Further analysis shows that not only do low signal patterns have low visibility, which was expected due to background but, contrary to expectations, very high signal frames show a similar reduction in visibility.

Figure~\ref{fig:setup}a schematically illustrates the experimental setup which was performed at the Single Particles, Biomolecules and clusters/Serial Femtosecond Crystallography (SPB/SFX) instrument~\cite{mancuso2019single} at the European X-ray free electron laser (EuXFEL). \SI{88}{\nano\meter} copper oxide nanocubes were aerosolized using an electrospray sample injector~\cite{electrospray} and directed into the focus of the EuXFEL beam, with a nominal focal spot size of $250\times$\SI{250}{\nano\meter\squared}. A \SI{20}{\micro\meter} thick nickel foil (K-edge at \SI{8332}{\electronvolt}~\cite{henke1993x}) was placed between the interaction region and the adaptive gain integrating pixel detector (AGIPD)~\cite{agipd} positioned \SI{715}{\milli\meter} away to preferentially absorb the elastic scattering and the K$\beta$ fluorescence while letting the K$\alpha$ photons through. The nickel foil had a \SI{10}{\milli\meter} by \SI{10}{\milli\meter} square aperture in the central region, which allowed the coherent elastic scattering to pass through. The spectrum of each of the \SI{9.1}{\kilo\electronvolt}, \SI{200}{\micro\joule} incoming X-ray pulses was also measured using a single-shot spectrometer~\cite{kujala2020hard}. Assuming the temporal spikes from the XFEL process were transform limited, the pulse duration was estimated using spectral autocorrelations to be \SI{2.8}{\femto\second} (details in End Matter).

Figure~\ref{fig:setup}b depicts part of an idealized detector frame where a few photons are recorded. The $\mathbf{q}$ vector joining two of the photons represents the wave vector difference in Eq~\ref{eq:g2_siegert}. For $N$ photons measured in a single exposure, there are $N^2$ photon pairs, contributing to the signal in Eq.~\ref{eq:g2_def} at different $\mathbf{q}$ values. However, not all pairs contribute to the signal, including cases where one of the photons is generated by elastic scattering or if both photons have different polarizations. The two photons can also be temporally or energetically separated. The effect of these non-contributing photon pairs combines to suppress the visibility factor $\nu$, since we expect the cross-correlation of non-interfering photons to be independent of their separation $\mathbf{q}$, not contributing to the second term.

\begin{figure}
    \centering
    \includegraphics[width=0.8\linewidth]{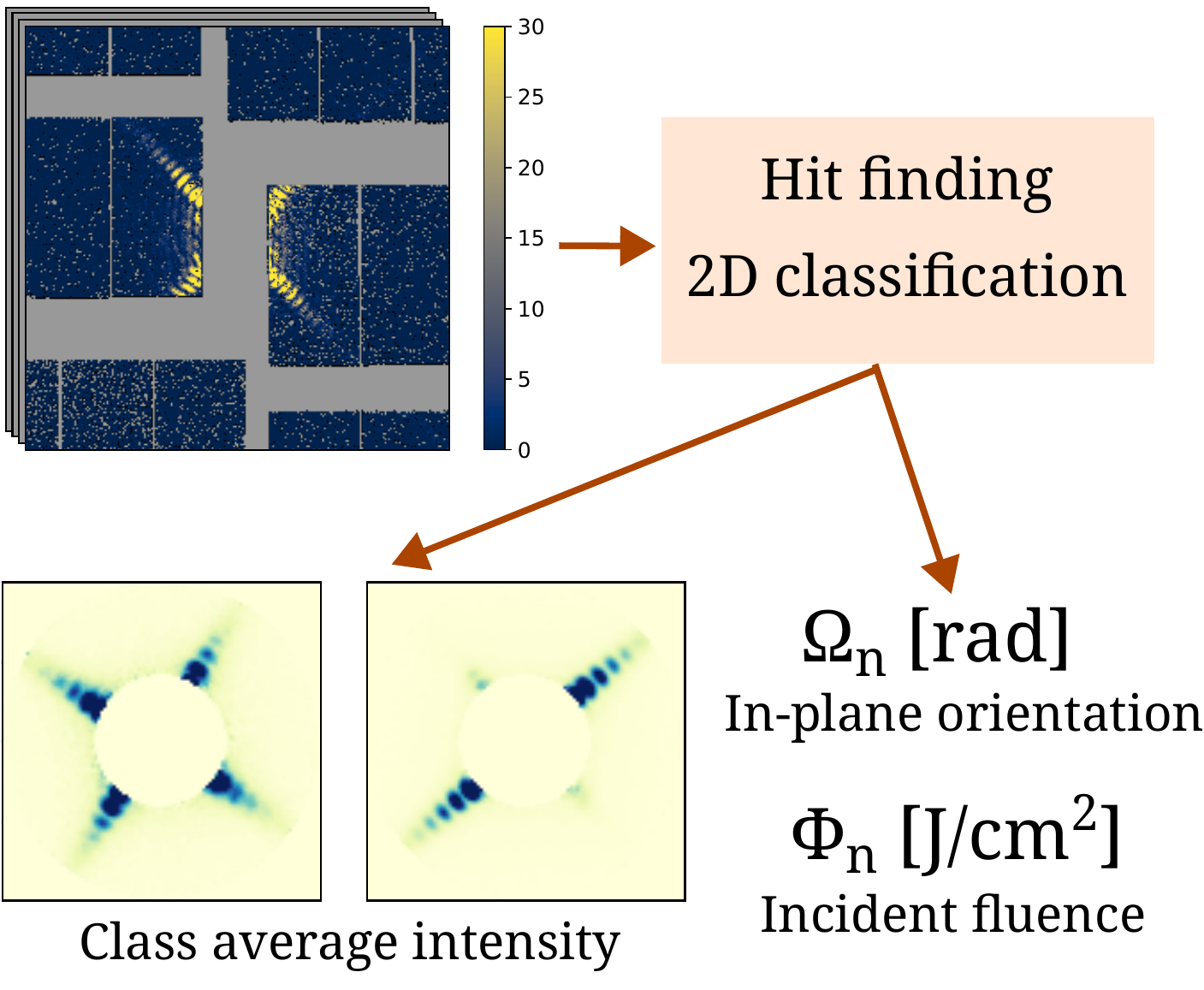}
    \caption{Single particle coherent diffractive imaging analysis. The central region of detector frames (top left) were analyzed to detect those where a particle was hit by the XFEL. 2D classification was then performed on these hits to yield class average coherent intensities as well as estimates of the incident fluence and in-plane orientation for each frame.}
    \label{fig:spi_analysis}
\end{figure}

\begin{figure}[t]
  \centering
  \includegraphics[width=0.8\columnwidth]{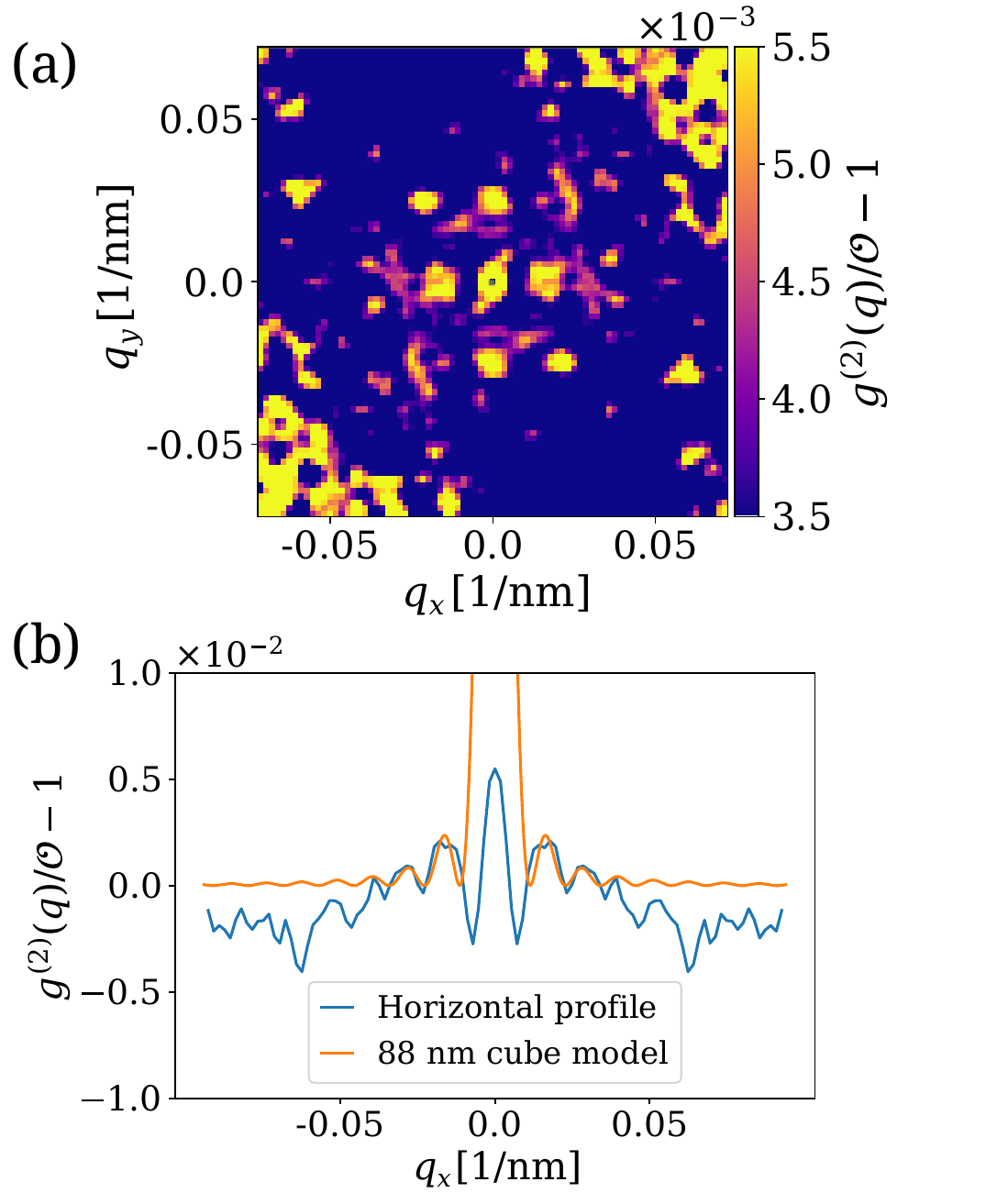} \\
  \caption{(a) Aligned $g^{(2)}(q_x, q_y)$ calculated from 1780 frames selected after 2D classification of the low-$q$ CDI signal. The single-frame $g^{(2)}(\mathbf{q})$ were rotated in-plane by the same amount required to align the CDI patterns. (b) Horizontal line profile $g^{(2)}(q_x, q_y=0)$ from (a), compared with the expected $|\mathcal{F}(\mathbf{q})|^2$ from an \SI{88}{\nano\meter} cube.}
  \label{fig:class_g2}
\end{figure}

Due to the sample delivery method, the sample in the X-ray focus is not the same for every X-ray pulse. Some of the pulses do not intersect any particle at all, and when they do, the orientation of the particle is random since the particles are tumbling. The small angle coherent diffraction detected through the window in the Ni filter was analyzed to find the patterns with diffraction from single nanocubes. In total, \num{316109} frames were detected with statistically significant signal over background. 2D classification of the elastic scattering region was performed using the EMC algorithm~\cite{loh2009reconstruction,ayyer20213d} as implemented in the \emph{Dragonfly} software~\cite{dragonfly} in two rounds. In each round, the code iteratively refined multiple maximum likelihood intensity models and in parallel, estimates of the in-plane orientation and relative incident fluence for each frame (Fig.~\ref{fig:spi_analysis}). In the first round, clusters and contaminants were filtered out, with the second round used to classify the different single nanocube patterns into groups of similar orientation and size. 

Since the particles can be intercepted anywhere from the center of the X-ray focus to its extremities, the incident fluences vary over multiple orders of magnitude, with a lower detectability threshold set by background scattering, primarily from the carrier gas. Altogether, \num{130985} single particle diffraction patterns were detected. The relative fluence factors from the classification were then converted to an absolute scale using tabulated coherent scattering cross sections and the fitted particle size. Note that this conversion assumes a fluence-independent coherent scattering cross-section, and experiments in the soft X-ray regime have shown an enhancement of the cross-section at higher fluences~\cite{kuschel2025non,ho2020role}.

The intensity correlation analysis was performed on the region of the detector shadowed by the nickel foil, which transmits only 0.88\% of the elastically scattered \SI{9.1}{\kilo\electronvolt} photons, while transmitting 43.5\% of the K-shell fluorescence from the copper nanocubes. Due to the relatively short distance between the interaction region and the detector, the outgoing wave-vectors $\mathbf{k}$ have an appreciable component along the beam direction which cannot be ignored i.e. the Ewald sphere curvature cannot be neglected. Thus, the wave-vector differences, $\mathbf{q}$, span a three-dimensional region~\cite{classen2017incoherent}. The two-point correlation function was calculated by counting the number of photon pairs with a given difference $\mathbf{q}$ in each frame and dividing by the 3D auto-correlation of the average frame, $\left\langle I(\mathbf{k}) \right\rangle$. If the only source of photons is a stable, Poisson-distributed background, these two quantities are identical, yielding $g^{(2)}(\mathbf{q}) = 1$ for all $\mathbf{q}$.

From the \textit{Dragonfly} class averages we selected and merged the $g^{(2)}$ signal of five similar classes, corresponding to 1780 frames with an average signal level of $\mu=4.5\times 10^{-3}$ photons/pixel/frame. The $q_x=0$ and $q_y=0$ lines were masked prior to alignment due to spurious correlations in detector noise. Figure~\ref{fig:class_g2}a shows an image averaged over 5 $q_z$ layers ($9.1\times10^{-3}$ 1/nm in width) and the horizontal line profile of the experimental data in Fig.~\ref{fig:class_g2}b aligns well with the expected Fourier amplitudes of an $\SI{88}{\nano\meter}$ cube model, with visible fringes up to a resolution of approximately $\SI{20}{\nano\meter}$. Note however, that the experimental profile also shows another feature, which is a smooth increase in the correlation as one approaches $\mathbf{q}=0$. This is a clue to alternative physics, which we will discuss below.

Analogous to standard time-domain HBT experiments for single photon emitters, where the degree of anti-bunching for $\Delta t = 0$ is retrieved~\cite{bourrellier2016bright,merino2015exciton}, we extrapolate the low-$\mathbf{q}$ $g^{(2)}$ signal to $\mathbf{q}=0$ to estimate the visibility factor $\nu$ in Eq~\ref{eq:g2_siegert} (see extrapolation details in the End Matter). This visibility has so far been defined as the difference between the signal at $\mathbf{q}=0$ and the overall offset $\mathcal{O}$ of the $g^{(2)}(\mathbf{q})$~\cite{inoue2019determination,trost2023imaging}. However, in our case, the average fluorescence signal on the detector, $\mu$ ($\langle I(\mathbf{k})\rangle$, measured in photons per pixel) varies dramatically from shot to shot, leading to a proportional increase of the $g^{(2)}$ signal at all $\mathbf{q}$
\begin{equation}
    g^{(2)} \propto 1 + \frac{\text{Var}(\mu)}{\langle \mu \rangle^2},
\end{equation}
where $\langle \mu \rangle$ denotes the average over many exposures. This increases both the constant offset in Eq.~\ref{eq:g2_siegert} as well as the $\mathbf{q}$-dependent modulation. To mitigate this effect, we now define the visibility as 
\begin{equation}
    \nu = \frac{g^{(2)}(\mathbf{q}=0)}{\mathcal{O}} - 1,
\end{equation}
where the offset $\mathcal{O}$ is the median $g^{(2)}$ value for all $\mathbf{q}$ (see validation from simulation in the End Matter).

\begin{figure}[t]
  \centering
  \includegraphics[width=\columnwidth]{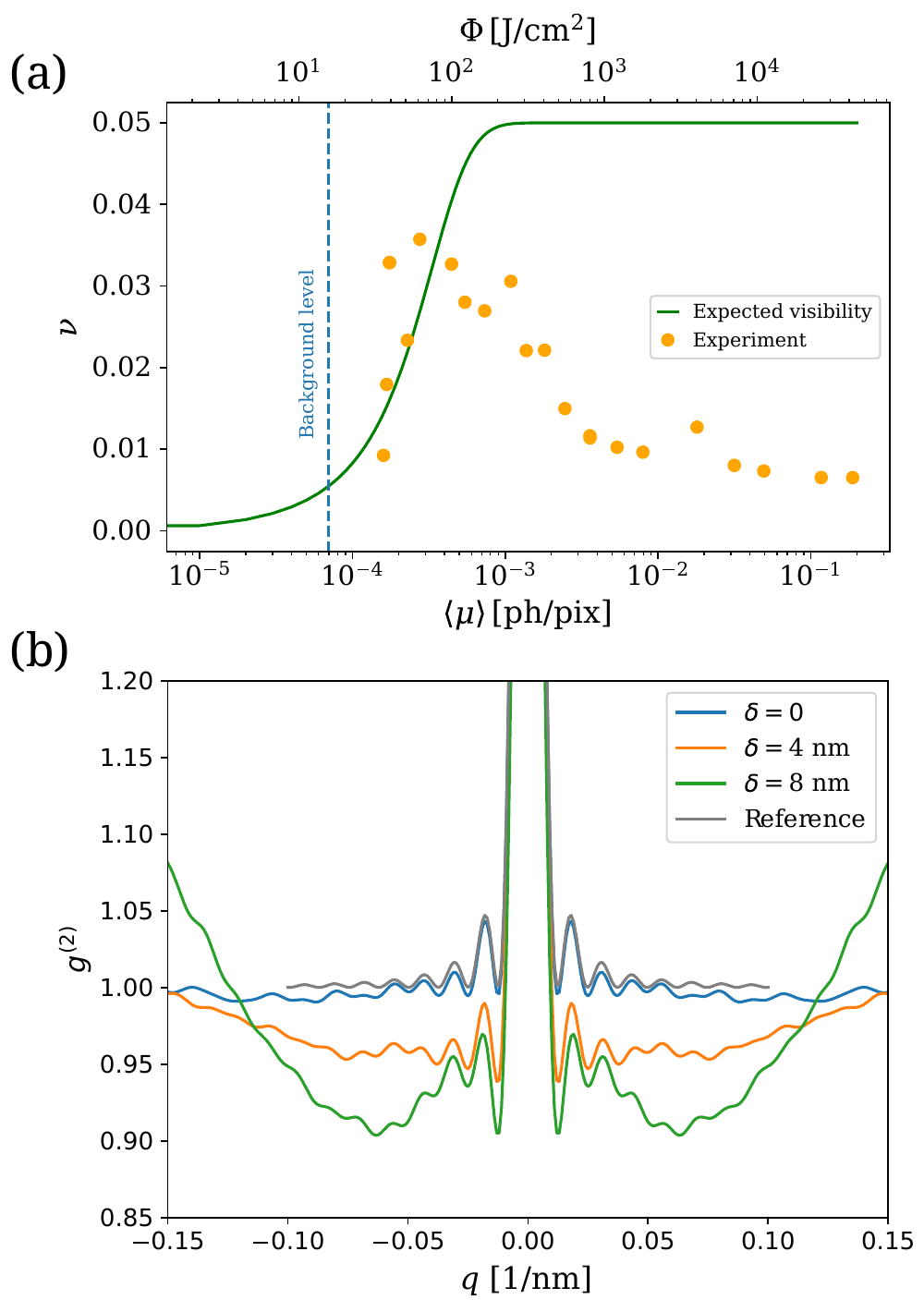}
  \caption{(a) Visibility $\nu$ as a function of $\langle\mu\rangle$ for subsets of frames. The green line shows the expected visibility for 20 temporal modes with the experimental background level (dashed blue line). (b) Simulated intensity correlation, $g^{(2)}$ for the case where the random initial phases are partially dependent on the values of the neighbors with a correlation length, $\delta$. Uncorrelated initial phases ($\delta=0$) match well with the reference plot from the Siegert relation.}
  \label{fig:vis}
\end{figure}

From the estimated pulse duration and photon statistics, we would expect the visibility to be 0.05~\cite{trost2023speckle,inoue2019determination}. However, the observed visibility in Fig.~\ref{fig:class_g2} was much lower. The single particle approach allows us to bin the experimental data into subsets of similar signal levels. Figure~\ref{fig:vis}a shows the visibility as a function of the average signal level, $\langle\mu\rangle$, for the experimental data. The green line shows the expected behavior for 20 temporal modes ($\nu=0.05$), alongside the constant measured elastic background from the carrier gas. The simulations show a reduction in the visibility for low $\mu$ due to the increasing contribution of the gas background, which is also reflected in the experimental data. But there is a large discrepancy at high $\langle\mu\rangle$, where instead of the visibility saturating to 0.05, we observe a drastic drop to $\nu = 0.007$, corresponding to approximately $M=140$ temporal modes.  This fall-off shows the same sigmoid-like behavior as the background-induced reduction at low $\langle\mu\rangle$, suggesting the appearance of a new source of background that does not contribute to the visibility. The ratio of the spontaneous fluorescence and this new source of background seems to have stabilized, leading to a lowered, but finite visibility.

We propose amplified spontaneous emission (ASE)~\cite{yoneda2015atomic,benediktovitch2020amplified} that fluctuates from shot to shot as a consistent explanation for these observations. In this phenomenon, rapid ionization at time scales shorter than the core-hole lifetime can lead to a population inversion. The emission of one K$\alpha$ photon then stimulates emission from other atoms in phase. These photons, at the K$\alpha$ photon energy, will be otherwise indistinguishable on the detector from the spontaneous fluorescence which contributes to the visibility. In foils much thicker than the focal spot size -- on which all prior experiments have been conducted -- ASE co-propagates with the XFEL due to the high aspect ratio of the illuminated volume (echoing the nuclear-pumped X-ray laser concept~\cite{melnikov2015lasers}). For cubic particles like those studied here, the spatial distribution of the ASE is less clear. Prior theoretical work has suggested ASE should have random coherence patches smaller than the full volume due to multiple initial relaxations, especially at lower fluences~\cite{chuchurka2024stochastic}. 

To further validate this, we simulate the intensity correlation signal in cases where the initial phases of the emitted photons are correlated with their neighbors with a finite correlation length, $\delta$ (as opposed to conventional IDI, where phases are assumed to be uncorrelated). Figure~\ref{fig:vis}b shows the simulated $g^{(2)}$ with $\nu=1$ for correlation lengths of 0, 4 and \SI{8}{\nano\meter} from an \SI{80}{\nano\meter} cubic particle. One can see the appearance of the smooth low-$\mathbf{q}$ features also observed in Fig.~\ref{fig:class_g2}b. No strong visibility reduction is observed, however, suggesting that this does not fully explain the observations.

In conclusion, we have presented a first demonstration of single-particle IDI at a hard X-ray FEL, with full single-particle classification and orientation from coherent diffraction and IDI analysis performed on the same pulses. We achieved element-specific imaging of nanocubes down to \SI{20}{\nano\meter} resolution, more than one order of magnitude better than previous IDI experiments~\cite{trost2023imaging,inoue2019determination}. At the same time the number of required shots could be reduced by four orders of magnitude, both due to the higher incident pulse energy and due to the fact that the smaller sample size allowed a much closer detector placement and collection of a larger fraction of the total emitted photons. 

We also experimentally observed a strong, non-monotonic reduction of IDI visibility at high fluence above $\SI{1e2}{\joule/\centi\meter^2}$, that we link to the emergence of amplified spontaneous emission under extreme illumination conditions. Thus, it appears that intense X-ray pulses alter the electronic structures of the particles in a manner that reduces the efficiency of IDI. This regime of strongly nonlinear absorption is investigated in detail in the companion Letter~\cite{cardoch2025nonlinear}, which provides complementary evidence for the fluence-dependent changes observed here. To overcome these limitations and to validate our hypothesis of ASE, we suggest further investigations using a spectrally-resolved version of IDI~\cite{wollweber2024nanoscale}, where the degree of coherence can be studied simultaneously for different emission energies.

\begin{acknowledgments}
\vspace{1em}
We acknowledge European XFEL in Schenefeld, Germany, for provision of X-ray free-electron laser beamtime at SPB/SFX SASE1 under proposal number 5476 and would like to thank the staff for their assistance. Data recorded for the experiment at the European XFEL are available at doi:10.22003/XFEL.EU-DATA-005476-00. This work is supported by the Cluster of Excellence `CUI: Advanced Imaging of Matter' of the Deutsche Forschungsgemeinschaft (DFG) - EXC 2056 - project ID 390715994. S.\ C.\ and N.\ T.\ acknowledge the Swedish Research Council (Grants No. 2019-03935 and No. 2023-03900) for financial support. C.\ C.\ acknowledges the Swedish Research Council (Grant No. 2018-00740), the Röntgen-Ångström Cluster (Grant No. 2019-03935), and the Helmholtz Association through the Center for Free-Electron Laser Science at DESY. S.K and J.v.Z acknowledge funding by the Deutsche Forschungsgemeinschaft within the TRR 306 QuCoLiMa (“Quantum Cooperativity of Light and Matter”), project-ID 429529648. This research was supported in part through the Maxwell computational resources operated at Deutsches Elektronen-Synchrotron DESY, Hamburg, Germany.
\end{acknowledgments}

\bibliography{lib}

\clearpage
{\centering \large{\textbf{End Matter}}}

\emph{Sample preparation---}
Copper(I) oxide ($\mathrm{Cu}_2\mathrm{O}$) nanocubes (NCs) were synthesized based on the protocol in \textcite{thoka}, which was upscaled 25-fold. Copper(II) sulfate pentahydrate (CuSO$_4$.5H$_2$O $\geq 99\%$), (+)-sodium L-ascorbate ($\geq 98\%$), and sodium dodecyl sulfate (SDS, $\geq 98.5\%$) were purchased from Sigma-Aldrich (USA). Sodium hydroxide (NaOH, $\geq 99\%$) was obtained from Roth (Germany). Ethanol (denat., $\geq 96\%$) was bought from VWR (USA). $\alpha$-Methoxypoly(ethylene glycol)-$\omega$-(11-mercaptoundecanoate) (PEGMUA:\SI{2}{\kilo\dalton} or \SI{5}{\kilo\dalton}) was synthesized as described by \textcite{schulz}. Ultrapure water (\SI{18.2}{\mega\ohm}, Millipore) was used for all procedures.

SDS ($\SI{1.44}{\gram}, \SI{5}{\milli\mole}$) was dissolved in $\SI{239}{\milli\liter}$ of ultrapure water and the solution was stirred for 5 min. Then $\mathrm{Cu}_2\mathrm{SO}_4 \cdot 5\mathrm{H}_2\mathrm{O}$ (\SI{2.5}{\milli\liter}, \SI{0.1}{\mole\per\liter}) and NaOH (\SI{1}{\milli\liter}, \SI{1.0}{\mole\per\liter}) were added under constant stirring. The solution turned light bluish after a few seconds which confirmed the formation of $\mathrm{Cu(OH)}_2$. Sodium ascorbate (\SI{7.5}{\milli\liter}, \SI{0.2}{\mole\per\liter}) was added and the mixture stirred for another 5 minutes. Subsequently, an undisturbed aging for 10 minutes was required for crystal growth and the solution turned bright orange. The particles were centrifuged at 9500g for 10 min and further washed (thrice for 10 minutes at 9500g) with a 1:1 volume ratio of water and ethanol. Finally, the $\mathrm{Cu}_2\mathrm{O}$ NCs were dissolved in $\SI{10}{\milli\liter}$ of ethanol and functionalized by adding PEGMUA (\SI{2}{\kilo\dalton} or \SI{5}{\kilo\dalton}, \SI{10}{\milli\liter}, \SI{1}{\milli\mole\per\liter}). The solution was stirred overnight followed by further purification (thrice for 10 minutes at 9500g). At the end, the solution was concentrated to a volume of $\SI{1}{\milli\liter}$. 

\begin{figure}
  \centering
  \includegraphics[width=0.6\columnwidth]{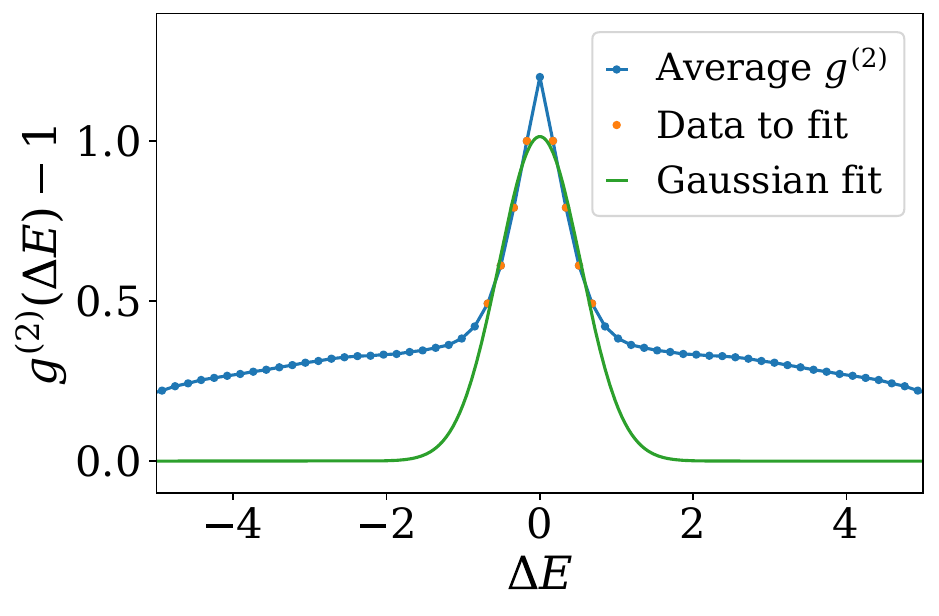}
  \caption{Average second order correlation function~$g^{(2)}(\Delta E)$. The full width at half maximum of the central feature is inversely proportional to the pulse duration.}
  \label{fig:hirex_g2}
\end{figure}

\emph{Pulse duration estimation---}
XFEL pulses produced through the self-amplified spontaneous emission (SASE) process have a spiky temporal profile, with individual spikes having durations of around \SI{200}{\atto\second} in the hard X-ray regime~\cite{yan2024terawatt}. In the spectral domain, this also yields spikes with an inverse relationship, where the spike width in energy is related to the total pulse duration and the overall spectral envelope is governed by the temporal spike duration. Since the location and spacing of the temporal spikes is random from pulse-to-pulse, one can calculate the spike width, and thus the pulse duration, by calculating the autocorrelation of the spectrum. This is shown in Fig.~\ref{fig:hirex_g2}, from which one can estimate the average pulse duration to be \SI{2.8}{\femto\second}.

\begin{figure}
  \centering
  \includegraphics[width=0.8\linewidth]{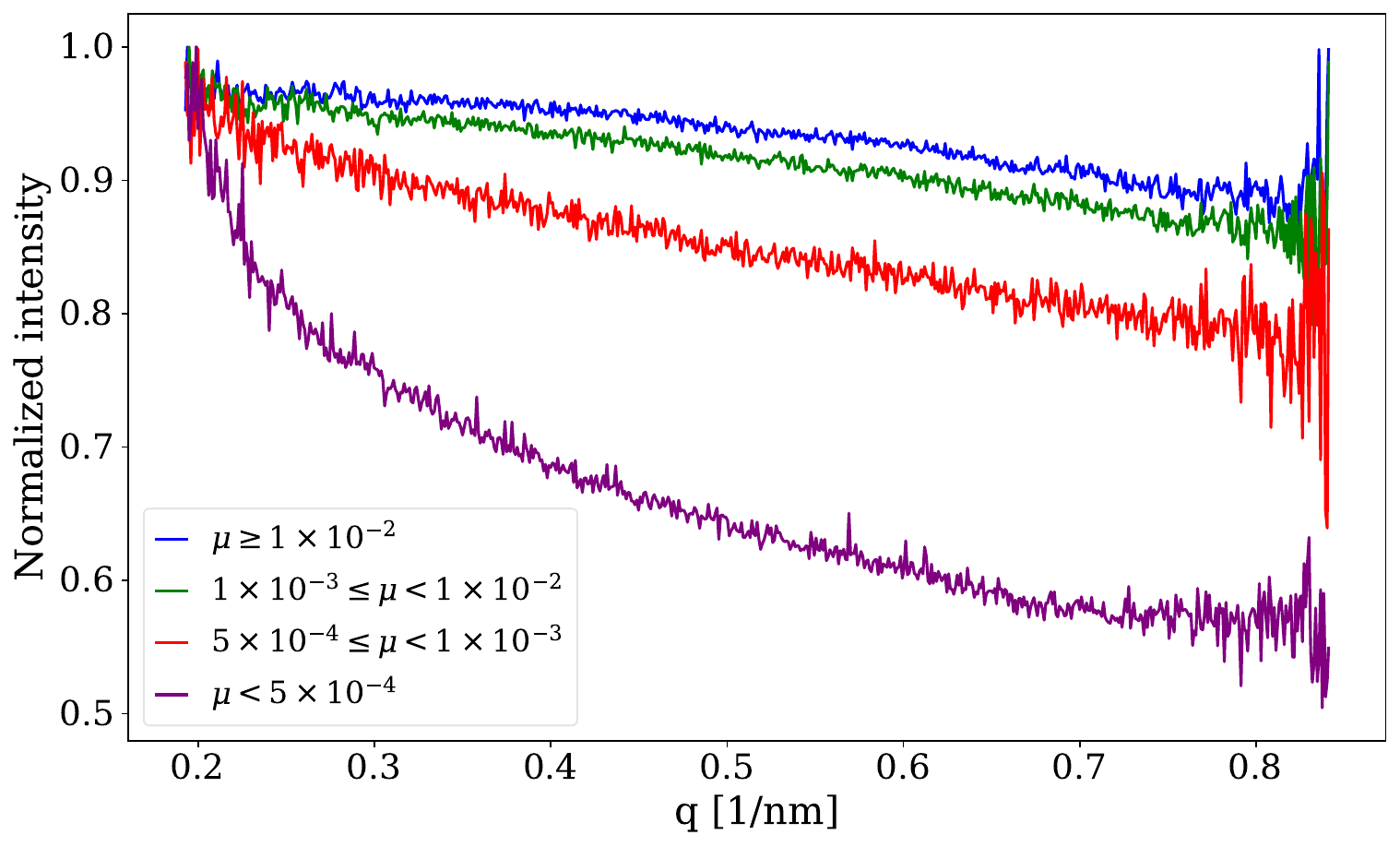}
  \caption{Normalized radial intensity profiles for subsets of patterns, binned by their signal level, $\mu$.}
  \label{fig:i_vs_q}
\end{figure}

\emph{Elastic scattering background---}
While the nickel filter blocked nearly all elastic scattering from the interaction region and further upstream, we still observe background from gas scattering downstream, which can reach the outer region of the detector through the square aperture. This was a stable background with an average signal of $\mu = 7\times10^{-5}$ photons/pixel/frame. One can see the effect of this background in the radial distribution of the fluorescent intensity for patterns with different $\langle\mu\rangle$ in Fig.~\ref{fig:i_vs_q}.

\begin{figure*}
\centering
    \includegraphics[width=0.65\textwidth]{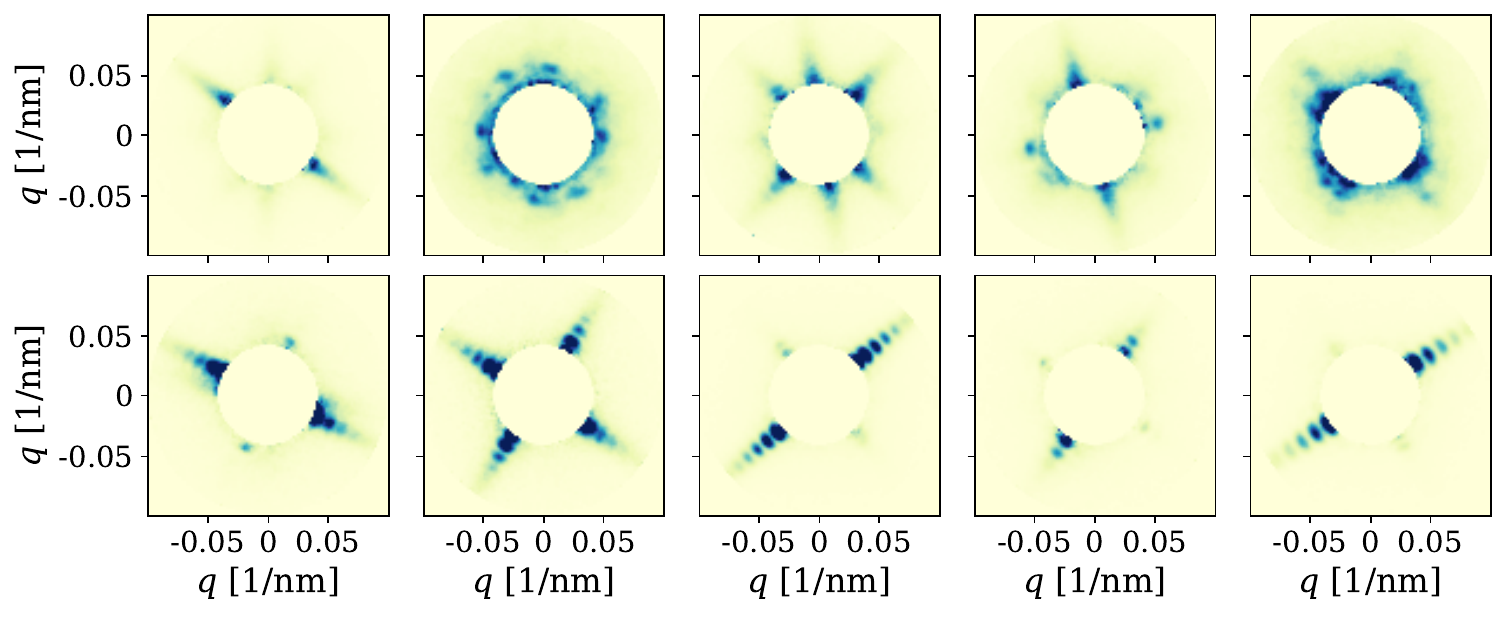}
\caption{\textit{Dragonfly} class averages. The top row displays five class averages that were rejected after the first round of classification. The frames belonging to the classes shown in the bottom row were used to calculate the $g^{(2)}$ signal shown in Fig~\ref{fig:class_g2}.}
\label{fig:classification}
\end{figure*}

\emph{Coherent diffraction classification---}
The diffraction patterns for short pulses were classified into 72 classes using the \textit{Dragonfly} software~~\cite{dragonfly, ayyer20213d}. Figure \ref{fig:classification} illustrates representative class averages: the top row includes those that were rejected after the initial classification round, while the bottom row shows the class averages of patterns that were selected for calculating the average $g^{(2)}$ signal presented in Fig~\ref{fig:class_g2}.

\emph{Absolute fluence estimation---}
We assumed that the elastically scattered intensity $I_\mathrm{el}$ scales linearly with the incident fluence $\Phi$ on the particles,
\begin{equation}
\label{eq:fluence}
     I_\text{el}(\mathbf{q}) = \Phi \cdot \left(2 \pi \cdot d^3 \cdot \Delta n \cdot \frac{1}{\lambda^2}\right)^2 \cdot \big| \mathcal{F}(\mathbf{q}) \big| ^2 \cdot \Omega(\mathbf{q}),
\end{equation}
where $d$ is the particle diameter, $\Delta n$ the refractive index of copper, $\lambda$ the incident photon wavelength and $\Omega(\mathbf{q})$ the integrated solid angle of the detector pixels where the elastic signal was being measured: 

\begin{equation}
    \Omega = \cos^3(\theta) \cdot \frac{s_p^2}{D^2}, 
\end{equation}
with the pixel at scattering angle $\theta$ of size $s_p$ and the detector distance $D=\SI{715}{\milli\meter}$.
$\big| \mathcal{F}(\mathbf{q}) \big|^2$ in Eq.~\ref{eq:fluence} are the normalized Fourier amplitudes of a cube model for the respective pixel at a scattering vector $\mathbf{q}$:

\begin{equation}
    \mathcal{F}(\mathbf{q}) = \frac{\sin(\pi \mathbf{q} d)}{\pi \mathbf{q} d}.
\end{equation}

From the \textit{Dragonfly} class averages we estimate an average particle size of $d=\SI{88}{\nano\meter}$. Hence, we can estimate the incident fluence on the particle from reference diffraction patterns using Eq.~\ref{eq:fluence}. During the classification process, \textit{Dragonfly} calculates the relative intensity scaling between different diffraction patterns in order to calculate a weighted class average. Using the calculated fluence for the reference diffraction patterns, we calculated the fluence for all diffraction patterns by applying the relative scaling factors. 


\begin{figure}
\centering
\includegraphics[width=0.8\columnwidth]{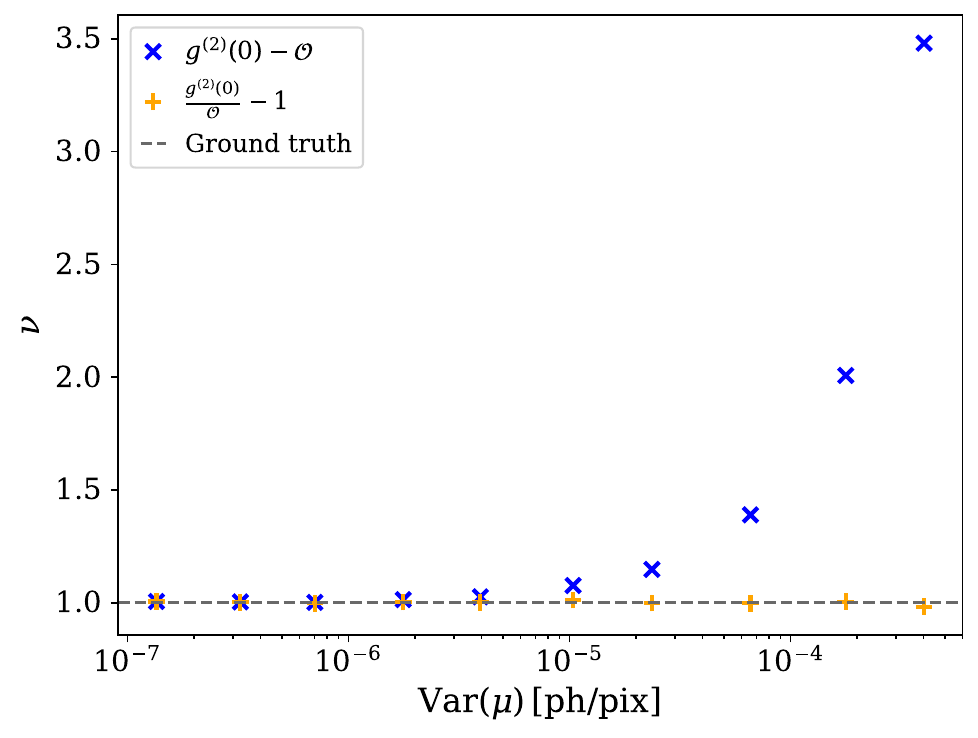}
\caption{Calculated visibility for datasets with different range of incident fluence fluctuations. The offset-subtraction method (blue) yields unphysical $\nu>1$.}
\label{fig:vis_fluct}
\end{figure}

\emph{Extrapolation of the visibility---}
The visibility at $\mathbf{q}=0$ cannot be measured directly as the experimental $g^{(2)}(\mathbf{q}=0)$ is not physically meaningful. Moreover, the $g^{(2)}$ signals exhibit artificially strong correlations for photons detected within the same pixel rows and columns. Therefore, we masked the voxels in the  $g^{(2)}$ where $q_x=0$ or $q_y=0$. To estimate the visibility at $\mathbf{q}=0$, we calculated the mean $g^{(2)}$ value for the inner four voxels $q_{vox_x},q_{vox_y} \in \{-1,1\}$. When applied to the simulations of a $\SI{88}{\nano\meter}$ particle, we retrieve a correction factor of $0.84$ from the mean deviation to the expected values ($\nu=1/M$).

\emph{Visibility and fluence fluctuations---}
One can calculate the effect of fluence fluctuations on $g^{(2)}(\mathbf{q})$ in the limit that the underlying true correlations $\langle I_1 \cdot I_2 \rangle$ are independent of the actual number of photons $N$. The average $g_n^{(2)}$ then becomes
\begin{align}
    g^{(2)}_{\text{n}}(\textbf{q}) &= \frac{\langle \langle NI(\textbf{k}) \cdot NI(\textbf{k}+\textbf{q}) \rangle \rangle_n}{\langle \langle NI(\textbf{k})\rangle^2\rangle_n} \notag \\[10pt]
    &= \frac{\langle N^2 \rangle_n}{\langle N \rangle_n^2} \cdot \frac{\langle I(\textbf{k}) \cdot I(\textbf{k}+\textbf{q}) \rangle}{\langle I(\textbf{k}) \rangle^2} \notag \\[10pt]
    &= \left(1 + \frac{Var(\mu)}{\langle \mu \rangle^2} \right) \cdot g^{(2)}_{\text{constant}}(\mathbf{q}) \label{eq:g2N}
\end{align}
Here, $g^{(2)}_{\text{constant}}(\mathbf{q})$ is the intensity correlation for constant fluence.  Figure~\ref{fig:vis_fluct} shows this prediction in simulations and the effect of different ways of accessing the normalized Fourier transform from the Siegert relation.

\emph{IDI with phase correlations---}
The simulated $g^{(2)}(q)$ profiles shown in Fig.~\ref{fig:vis}b were calculated in one dimension by propagating waves with random, but correlated, initial phases to the far field. For each exposure, the intensities were calculated by integrating 161 emitters uniformly spaced along the length of the \SI{80}{\nano\meter} particle,
\begin{equation}
    I(q) = \left|\sum_n e^{i(qx_n + \phi_n)}\right|^2
\end{equation}
where $x_n$ is the position of the $n$-th emitter. In conventional IDI, the initial phases, $\phi_n$, are assumed to be random and uncorrelated. In this case, these phases were drawn from a joint normal distribution with a covariance matrix which decayed as a Gaussian function of the distance between two emitters,
\begin{equation}
    C_{nn'} \propto \exp\left[-\frac{(x_n-x_{n'})^2}{2 \delta^2}\right]
\end{equation}
$10^4$ exposures were generated and no Poisson noise was imposed.

\end{document}